\documentstyle[prl,aps,psfig,preprint]{revtex}
\draft
\global\firstfigfalse
\tightenlines
\newcommand{\nc}{\newcommand}\newcommand{\rnc}{\renewcommand}
\rnc{\/}{\over}\nc{\<}{\langle}\rnc{\>}{\rangle}\rnc{\/}{\over}
\rnc{\(}{\left (}\rnc{\)}{\right )}\rnc{\[}{\left [}\rnc{\]}{\right ]}
\nc{\Tr}{\text{tr}}\rnc{\i}{\text{i}}\nc{\e}{\text{e}}
\nc{\nq}{{\cal N}}
\begin{document}
\title{Spectral Statistics for Quantum Graphs:\\ Periodic Orbits and
Combinatorics}
\author{Holger Schanz$^\dag$ and Uzy Smilansky$^\ddag$\\[5mm]}
\address {$^\dag$Max-Planck-Institut f\"ur Str\"omungsforschung,\\ 37073
G\"ottingen, Germany\\[2mm]}
\address{$^\ddag${Department of Physics of Complex Systems,} \\
{The Weizmann Institute of Science, Rehovot 76100, Israel}\\[5mm]}
\date{April 25, 1999}
\maketitle
\begin{center}\mbox{}\\
To be published in the\\
{\it Proceedings of the Australian Summer School on Quantum
Chaos and Mesoscopics}\\
Canberra, Australia, January 1999
\end{center}
\begin{abstract}
We consider the Schr\"odinger operator on graphs and study the spectral
statistics of a unitary operator which represents the quantum evolution, or a
quantum map on the graph. This operator is the quantum analogue of the
classical evolution operator of the corresponding classical dynamics on the
same graph. We derive a trace formula, which expresses the spectral density of
the quantum  operator in terms of periodic  orbits on the graph, and show that
one can reduce the computation of the  two-point spectral correlation
function to a well defined combinatorial problem. We  illustrate this
approach by considering  an ensemble of
simple graphs. We prove by a direct computation that the two-point correlation
function  coincides with the CUE expression for $2\times2$ matrices. We
derive the same result using the periodic orbit approach in its combinatorial
guise.  This involves the use of advanced combinatorial techniques
which we explain.
\end{abstract}
\pacs{05.45.+b, 03.65.Sq}
\section {{\bf Introduction} \label{introduction}}
We have recently shown \cite{KS97,KS99} that the Schr\"odinger operator on
graphs provides a useful paradigm for the study of spectral statistics and
their relations to periodic orbit theory.  In particular, the universal
features which are observed in quantum systems whose classical counterpart is
chaotic, appear also in the spectra of quantum graphs. This observation was
substantiated by several numerical studies. The relevance to quantum chaology
was established by identifying the underlying {\it mixing} classical evolution
on the graphs, which provides the stability coefficients and actions of
periodic orbits in whose terms an exact trace formula can be written
\cite{R83,KS97,KS99}.

In spite of the large amount of effort invested in the past fifteen years
\cite{berry,keatbog}, we have only a limited understanding of the reasons for
the universality of spectral statistics in systems whose classical dynamics is
chaotic. The main stumbling block is the lack of understanding of the
intricate and delicate interference between the contributions of
(exponentially many) periodic orbits. This genuinely quantum quantity, (also
known as the ``off-diagonal" contribution), is the subject of several
researches, which address it from various points of view \cite
{keatbog,ADDKKSS93,CPS98,Miller97,Agam95}.  The present contribution attempts
to illuminate this issue from yet another angle, and we harness for this
purpose quantum graphs and combinatorics.

Our material is presented in the following way. We shall start by defining the
quantum dynamics on the graph in terms of a quantum map.  This map will be
represented by a unitary matrix, which is the quantum analogue of the
classical Frobenius-Perron operator of the properly defined classical dynamics
on the graph. The spectrum of the quantum operator is on the unit circle, and
its statistics is the main object of the present work.  After defining the
two-point correlation function of interest, we shall write it down in terms of
periodic orbits and discuss the combinatorial problem which should be
addressed in order to obtain a complete expression which includes the
``off-diagonal" contribution. Since the RMT is known to reproduce the
two-point correlation function for generic graphs, we propose that the RMT
expression could be obtained from a combinatorial theory, perhaps as the
leading term in an asymptotic expansion. For one particular example we show
that this is indeed the case in the last section. There we construct an
ensemble of simple graphs with non-trivial spectral statistics, which can be
solved in two independent ways. The direct way yields the statistics of RMT
for the $2\times 2$ circular unitary ensemble (CUE). The corresponding periodic
orbit calculation is converted into a combinatorial problem, which is solved by
proving a previously unknown combinatorial identity.

\section {{\bf The Quantum Scattering Map and its Classical Analogue}
\label{maps}}
\subsection{General Definitions for Quantum Graphs}
We shall start with a few general definitions.  Graphs consist of $V$ {\it
vertices} connected by $B$ {\it bonds} (or {\it edges}). The {\it valency}
$v_{i}$ of a vertex $i$ is the number of bonds meeting at that vertex.
Associated to every graph is its {\it connectivity (adjacency) matrix}
$C_{i,j}$.  It is a square matrix of size $V$ whose matrix elements $C_{i,j}$
are given in the following way
\begin{eqnarray}
C_{i,j}=C_{j,i}=\left\{
\begin{array}{l}
1\qquad\text{if}\  i,j\  \text{ are connected} \\
0\qquad\text{otherwise}
\end{array}\right\} 
\qquad(i,j=1,\dots,V)\,.
\label{cmat}
\end{eqnarray}
The valency of a vertex is given in terms of the connectivity matrix, by $v_i=
\sum_{j=1}^V C_{i,j}$ and the total number of bonds is $B= {1\over
2}\sum_{i,j=1}^VC_{i,j}$.

When the vertices $i$ and $j$ are connected, we shall assume that the
connection is achieved by a single bond, such that multiple bonds are
excluded.  We denote the connecting bond by $b=[i,j]$.  Note that the notation
$[i,j]$ will be used whenever we do not need to specify the {\it direction} on
the bond.  Hence $[i,j]=[j,i]$.  {\it Directed bonds} will be denoted by
$(i,j)$, and we shall always use the convention that the bond is directed from
the first index to the second one. To each bond $[i,j]$ we assign a length
$L_{[i,j]} = L_{(i,j)} = L_{(j,i)}$. In most applications we would avoid
non-generic degeneracies by assuming that the $L_{[i,j]}$ are {\it rationally
independent}.  The mean length is defined by $\left \langle L \right \rangle
\equiv {1\over B}\sum_{b=1}^B L_b$.

For the quantum description we assign to each bond $b=[i,j]$ a coordinate
$x_b$ which measures distances along the bond.  We may use $x_{(i,j)}$ which
is defined to take the value $0$ at the vertex $i$ and the value $L_{(i,j)}
\equiv L_{(j,i)}$ at the vertex $j$. We can also use $x_{(j,i)}$ which
vanishes at $j$ and takes the value $L_{(i,j)}$ at $i$.

The wave function $\Psi$ is a $B-$component vector and will be written as
$(\Psi_{b_1}(x_{b_1})$, $\Psi_{b_2}(x_{b_2}),\dots$, $\Psi_{b_B}(x_{b_B}))^T$
where the set $\{b_i\}_{i=1}^B$ consists of all the $B$ distinct bonds on the
graph. We will call $\Psi_b(x_b)$ the component of $\Psi$ on the bond $b$. The
bond coordinates $x_b$ were defined above. When there is no danger of
confusion, we shall use the shorthand notation $\Psi_b(x)$ for $\Psi_b(x_b)$
and it is understood that $x$ is the coordinate on the bond $b$ to which the
component $\Psi_b$ refers.

The Schr\"{o}dinger equation is defined on the graph in the following way
\cite{A83,A94} (see also \cite{KS99} for an extensive list of references on
the subject): On each bond $b$, the component $\Psi_b$ of the total wave
function $\Psi$ is a solution of the one-dimensional equation
\begin{eqnarray}
\left(-\i\;{\text{d}/ \text{d}x_{(i,j)}}
-A_{(i,j)}\right)^2\Psi_b(x_{(i,j)})=k^2\Psi_b(x_{(i,j)})
\qquad(b=[i,j])\,.
\label{schrodinger}
\end{eqnarray}
We included a ``magnetic vector potential" $A_{(i,j)}$, with $A_{(i,j)}=
-A_{(j,i)}$ which breaks time-reversal symmetry.

On each of the bonds, the general solution of (\ref {schrodinger}) is a
superposition of two counter-propagating waves
\begin {eqnarray}
\psi_{(i,j)}(x_{(i,j)})   =   \exp
\left (\i\[ kx_{(i,j)} +
A_{(i,j)}x_{(i,j)}\]\right ) \nonumber
\\
\psi_{(j,i)}(x_{(j,i)})  =  \exp \left (\i\[ kx_{(j,i)}
+
A_{(j,i)}x_{(j,i)}\]\right )\,.
\label{counterprop}
\end{eqnarray} 
Note that the above functions are normalised to have an amplitude
$1$ at the points from which they ``emerge", namely, $\psi_{(i,j)}=1$ at the
vertex $i$ and $\psi_{(j,i)}=1$ at the vertex $j$.  The Hilbert space of the
solutions of (\ref {schrodinger}) is spanned by the set of functions defined
above, such that for all $b=[i,j]$
\begin{equation}
\Psi_b  = a_{(i,j)} \psi_{(i,j)}(x_{(i,j)}) +
a_{(j,i)} \psi_{(j,i)}(x_{(j,i)})\,.
\label {aijdef}
\end{equation}
Thus, the yet undetermined coefficients $a_{(i,j)}$ form a $2B$-dimensional
vector of complex numbers, which uniquely determines an element in the Hilbert
space of solutions. This space corresponds to ``free wave" solutions since we
did not yet impose any conditions which the solutions of (\ref {schrodinger})
have to satisfy at the vertices.
\subsection{The Quantum Scattering Map}
The {\em quantum scattering map} is a unitary transformation acting in the
space of free waves, and it
is defined as follows.

In a first step, we prescribe at each vertex $i=1,\dots,V$ a {\it vertex
scattering matrix} which is a unitary matrix of dimension $v_i$. The vertex
scattering matrices may be $k$ dependent and they are denoted by
$\sigma_{l,m}^{(i)} (k)$, where the indices $l,m$ take the values of the
vertices which are connected to $i$, that is, $C_{i,l}=C_{i,m}=1$. The vertex
scattering matrix is a property which is attributed to the vertex under
consideration. It can either be derived from appropriate boundary conditions
as in \cite {KS97,KS99}, or, it can be constructed to model other physical
situations.  The important property of $\sigma_{l,m}^{(i)} (k)$ in the present
context is, that any wave which is {\it incoming} to the vertex $i$ from the
bonds $(l,i)$, and which has an amplitude $1$ at the vertex, is scattered and
forms {\it outgoing} waves in the bonds $(i,m)$ with amplitudes
$\sigma_{l,m}^{(i)} (k)$.

Now, the quantum scattering map is represented by its effect on the
$2B$-dimensional vector of coefficients ${\bf a} = \left \{a_{(i,j)}\right \}
$, namely, $\bf a$ is mapped to $\bf a'$ with components
\begin{equation}
a'_{ b'} = \sum _{b=1}^{2B}a_{b}
S_{B_{{b,b'}}}\,,
\label{QSM}
\end{equation}
where $b$ and $b'$ run over all directed bonds, and if we denote $b=(i,j)$
and $b'= (l,m)$
\begin{equation}
 {S_B}_{(i,j),(l,m)}(k) = \delta _{j,l}
 {\rm
e}^{\i L_{(i,j)}(k+A_{(i,j)})}\sigma_{i,m}^{(j)} (k)\,.
\label{S_Bdef}
\end{equation}
The effect of $S_B$ on a wave function can be intuitively understood as
follows. The coefficient $a_{(i,j)}$ is the (complex) amplitude of the wave
which emerges from the vertex $i$ and propagates to the vertex $j$.  Once it
reaches the vertex $j$, it has accumulated a phase ${\rm e}^{i
L_{(i,j)}(k+A_{(i,j)})}$ and it scatters into the bonds which emanate from $j$
with an amplitude given by the appropriate vertex scattering matrix.  The new
amplitude $a'_{(l=j,m)}$ consists of the superposition of all the amplitudes
contributed by waves which impinge on the vertex $l=j$ and then scatter. The
name ``quantum scattering" map is justified by this intuitive picture.

The resulting matrix $S_B$ is a $2B \times 2B$ unitary matrix. The unitarity
follows simply from the unitarity of the vertex scattering matrices, and from
the fact that $S_B$ has non-vanishing entries between connected directed
bonds: the incoming bond aims at the vertex from which the outgoing bond
emerges.  The unitarity of $S_B$ implies that its spectrum is restricted to
the unit circle. In this paper we shall mainly be concerned with the spectral
statistics of the eigenphases, and their relation to the underlying classical
dynamics on the graph. The spectral statistics will be discussed in the next
chapter. We shall use the remaining part of the present chapter to clarify two
important issues. We shall first show how one can use the quantum scattering
map to construct the space of solutions of the Schr\"odinger operator on the
graph with boundary conditions. Then, we shall introduce the classical
dynamics which corresponds to the scattering map.

 To define the space of ``bound states" on the graph, one has to restrict the
space of wave functions by imposing appropriate boundary conditions on the
vertices.  The boundary conditions guarantee that the resulting Schr\"odinger
operator is self-adjoint. In \cite {KS97,KS99}, we described and used one
particular set of boundary conditions, which ensure continuity (uniqueness)
and current conservation.  Here we shall use a slight generalisation, which
matches well with the spirit of the present article. We shall impose the
boundary conditions in terms of a consistency requirement that the
coefficients $a_{(i,j)}$ have to obey. Namely, we require that the
wave function (\ref {aijdef}) is {\it stationary} under the action of the
quantum scattering map. In other words, the vector {\bf a} must be an
eigenvector of $S_B(k)$ with a unit eigenvalue.
(see also \cite{Rochus}). This requirement can be
fulfilled when
\begin{equation}
\det (I-S_B(k)) =0\,.
\label {secular}
\end{equation}
In \cite {KS97,KS99} we have actually derived (\ref {secular}), for the
particular case in which the vertex scattering matrices where computed form a
particular set of vertex boundary conditions which impose continuity and
current conservation on the vertices. The resulting vertex scattering matrices
read
\begin{equation}
\sigma _{j,j^{\prime }}^{(i)}=\left( -\delta _{j,j^{\prime }}+{\frac{
(1+\e^{-\i\omega _i})}{v_i}}\right) C_{i,j}C_{i,j^{\prime
}},\medskip\
\,\,\,\omega _i=2\arctan \frac{\lambda _i}{v_ik}\,.
\label{smatrix}
\end{equation}
Here, $0\le \lambda_i \le \infty$ are arbitrary constants. The ``Dirichlet"
(``Neumann") boundary conditions correspond to $ \lambda_i =\infty \ \ (0)$,
respectively. The Dirichlet case implies total reflection at the vertex,
$\sigma_{j,j^{\prime }} ^{(i)}= -\delta _{j,j^{\prime }}$. For the Neumann
boundary condition we have $\sigma _{j,j^{\prime }}^{(i)}=-\delta _{j,j^{\prime
}}+2/{v_i}$ which is independent of $k$. For any intermediate boundary
condition, the scattering matrix approaches the Neumann expression as $k
\rightarrow \infty$.  Note that in all non-trivial cases ($v_i> 2$),
back-scattering ($j =j^{\prime}$) is singled out both in sign and in
magnitude: $\sigma _{j,j }^{(i)}$ has always a negative real part, and the
reflection probability $|\sigma _{j,j }^{(i)}|^2 $ approaches $1$ as the
valency $v_i$ increases. One can easily check that $\sigma ^{(i)}$ is a
symmetric unitary matrix, ensuring flux conservation and time reversal
symmetry at the vertex.  For Neumann boundary conditions $\sigma ^{(i)}$ is a
real orthogonal matrix.

The spectral theory of the Schr\"odinger operators on graphs can be developed
using (\ref {secular}) as the starting point. In particular, the corresponding
trace formula \cite {R83} can naturally be derived, and related to the
underlying classical dynamics \cite {KS97,KS99}.  Here, we shall study the
quantum scattering map on its own right, without a particular reference to its
r\^ole in the construction of the spectrum.  We shall consider the ensemble of
unitary, $2B\times 2B$ matrices $S_B(k)$, where $k$ is allowed to vary in a
certain interval to be specified later. Our main concern will be the
statistical properties of the eigenvalues of $S_B$. This will be explained in
the next chapter.

\subsection{The Classical Scattering Map}
The last point to be introduced and discussed in the present chapter is the
classical dynamics on the graph and the corresponding scattering map.

We consider a classical particle which moves freely as long as it is on a
bond.  The vertices are singular points, and it is not possible to write down
the analogue of Newton's equations at the vertices. Instead, one can employ a
Liouvillian approach based on the study of the evolution of phase-space
densities.  This phase-space description will be constructed on a Poincar\'{e}
section which is defined in the following way. Crossing of the section is
registered as the particle encounters a vertex, thus the ``coordinate" on the
section is the vertex label. The corresponding ``momentum" is the direction in
which the particle moves when it emerges from the vertex.  This is completely
specified by the label of the next vertex to be encountered.  In other words,
\begin{equation}
\left\{
\begin{array}{c}
{\rm position} \\
{\rm
momentum}
\end{array}
\right\} \Longleftrightarrow
\left\{
\begin{array}{c}
{\rm vertex}\text{ }{\rm index} \\
{\rm
next}\text{ }{\rm index}
\end{array}
\right\}\,.
\end{equation}
The set of all possible vertices and directions is equivalent to the set of
$2B$ directed bonds. The evolution on this Poincar\'{e} section is well
defined once we postulate the transition probabilities $P_{j\to
j^{\prime}}^{(i)}$ between the directed bonds $b=\{j,i\}$ and $b^{\prime
}=\{i,j^{\prime }\}$.  To make the connection with the quantum description, we
adopt the quantum transition probabilities, expressed as the absolute squares
of the $S_B$ matrix elements
\begin{equation}
P_{j\to j^{\prime }}^{(i)}=\left|
\sigma_{j,j^{\prime
}}^{(i)}(k)\right|
^2\,.
\label{cl1}
\end{equation}
When the vertex scattering matrices are constructed from the standard matching
conditions on the vertices (\ref {smatrix}), we get the explicit expression
\begin{equation}
P_{j\to
j^{\prime }}^{(i)} =
\left| -\delta _{j,j^{\prime }}+{\frac{(1+\e^{-\i\omega_i})}{v_i}}\right|^2  \,.
\label{cl11}
\end{equation}
For the two extreme cases corresponding to Neumann and Dirichlet boundary
conditions this results in
\begin{eqnarray}
P_{j\to j^{\prime }}^{(i)} &=&
\left\{
\begin{array}{ll}
\left( -\delta _{j,j^{\prime }}+{2/{v_i}}\right) ^2 & \text{Neumann} \cr
\delta _{j,j^{\prime}} & \text{Dirichlet}
\end{array}
\right\}\,.
\end{eqnarray}
The transition probability $P_{j\to j^{\prime }}^{(i)}$ for the
Dirichlet case admits the following physical interpretation. The particle
is confined to the bond where it started and thus the phase space is divided
into non-overlapping ergodic components ($\approx$ ``tori''). For all other
boundary conditions the graph is dynamically connected.

The classical Frobenius-Perron evolution operator is a $2B\times 2B$ matrix
whose elements $U_{b,b^{\prime }}$ are the classical transition probabilities
between the bonds $b,b^{\prime }$
\begin{equation}
U_{ij,nm}=\delta_{j,n} P^{(j)}_{i\to m}\,.
\label{cl3}
\end{equation}
$U$ does not involve any metric information on the graph, and for Dirichlet or
Neumann boundary conditions $U$ is independent of $k$. This operator is the
classical analogue of the quantum scattering matrix $S_B$. Usually, one
``quantises" the classical operator to generate the quantum analogue. For
graphs the process is reversed, and the classical evolution is derived from
the more fundamental quantum dynamics.

 Let $\rho _b(t), \   b=1,\dots, 2B$ denote  the distribution of probabilities
to occupy the directed bonds  at the (topological) time $t$. This
distribution will evolve after the first return  to the Poincar\'e section
according to
\begin{equation}
\rho _b(t+1)=\sum_{b^{\prime }}U_{b,b^{\prime }}\rho _{b^{\prime }}(t)\,.
\label{master}
\end{equation}
This is a Markovian master equation which governs the evolution of the
classical probability distribution. The unitarity of the graph scattering
matrix $S_B$ guarantees $\sum_{b=1}^{2B}U_{b,b^{\prime }}=1$ and $0\leq
U_{b,b^{\prime }}\leq 1$, such that the probability that the particle is on any
of the bonds is conserved during the evolution. The spectrum of $U$ is
restricted to the unit circle and its interior, and $\nu_1 = 1$ is always an
eigenvalue with the corresponding eigenvector $|1\rangle = \frac 1{2B} \left(
1,1,...,1\right) ^T$. In most cases, the eigenvalue $1$ is the only eigenvalue
on the unit circle. Then, the evolution is ergodic since any initial density
will evolve to the eigenvector $|1\rangle $ which corresponds to a uniform
distribution (equilibrium).
\begin{equation}
\rho (t)\ \overrightarrow{\scriptstyle t{\rightarrow \infty }}\ |1\rangle\,.
\label{epart}
\end{equation}
The mixing rate $-\ln \left| \nu _2\right|$ at which equilibrium is approached
is determined by the gap between the next largest eigenvalue $\nu_2$ and $1$.
This is characteristic of a classically mixing system.

However, there are some non-generic cases such as, e.g., bipartite graphs when
$-1$ belongs to the spectrum. In this case the asymptotic distribution is not
stationary. Nevertheless an equivalent description is possible for bipartite
graphs when $U$ is replaced by $U^2$ which has then two uncoupled blocks of
dimension $B$. The example that we are going to discuss in the last section
will be of this type.

Periodic orbits on the graph will play an important r\^ole in the sequel and
we define them in the following way.  An {\it orbit} on the graph is an
itinerary (finite or infinite) of successively connected {\it directed} bonds
$\{i_1,i_2\}, \{i_2,i_3\},\dots $  For graphs without loops or multiple
bonds this is uniquely defined by the sequence of vertices $i_1,i_2, \dots$
with $i_m \in [1,V]$ and $C_{i_m,i_{m+1}} =1$ for all $m$.  An orbit is {\it
periodic} with period $n$ if for all $k$, $ (i_{n+k},i_{n+k+1})
=(i_k,i_{k+1})$. The {\it code} of a periodic orbit of period $n$ is the
sequence of $n$ vertices $i_1,\dots,i_n$ and the orbit consists of the bonds
$(i_m,i_{m+1})$ (with the identification $i_{m+n} \equiv i_{m}$). In this way,
any cyclic permutation of the code defines the same periodic orbit.

The  periodic orbits (PO's) can be classified in the following way:
\begin{itemize}
\item {\it Irreducible periodic orbits} - PO's which do not intersect
themselves such that any vertex label in the code can appear at most
once. Since the graphs are finite, the maximum period of irreducible PO's is
$V$. To each irreducible PO corresponds its time reversed partner whose code
is read in the reverse order. The only PO's which are both irreducible and
conjugate to itself under time reversal are the PO's of period 2.
\item
{\it Reducible periodic orbits} - PO's whose code is constructed by inserting
the code of any number of irreducible PO's at any position which is consistent
with the connectivity matrix. All the PO's of period $n >V$ are reducible.
\item {\it Primitive periodic orbits} -  PO's whose code
cannot be written down
as a repetition of a shorter code.
\end{itemize}
We introduced above the concept of orbits on the graph as strings of vertex
labels whose ordering obeys the required connectivity.  This is a finite
coding which is governed by a Markovian grammar provided by the connectivity
matrix. In this sense, the symbolic dynamics on the graph is Bernoulli. This
property adds another piece of evidence to the assertion that the dynamics on
the graph is chaotic. In particular, one can obtain the topological entropy
$\Gamma$ from the symbolic code. Using the relation
\begin {equation}
\Gamma = \lim_{n\to \infty} {1\over n} \log{\rm
tr} (C^n)
\end{equation}
one gets $\Gamma = \log \bar v$, where $\bar v$ is the mean valency.

Of prime importance in the discussion of the relation between the classical
and the quantum dynamics are the traces $u_n={\rm tr}(U^n)$ which are
interpreted as the mean classical probability to perform $n$-periodic motion.
Using the definition (\ref {cl3}) one can write the expression for $u_n$ as a
sum over contributions of $n$-periodic orbits
\begin{equation}
u_n=\sum_{p\in {\cal P}_n} n_p   \exp (-r \gamma_p
n_p ) \,,
\label{classicalsum1}
\end{equation}
where the sum is over the set ${\cal P}_n$ of primitive PO's whose period
$n_p$ is a divisor of $n$, with $r=n/n_p$. To each primitive orbit one can
assign a {\it stability factor} $\exp (-\gamma_p n_p ) $ which is accumulated
as a product of the transition probabilities as the trajectory traverses its
successive vertices:
\begin{equation}
 \exp (-\gamma_p n_p)  \equiv\prod _{j=1}^{n_p}
P^{(i_j)}_{i_{j-1}\to i_{j+1}}\,.
\label{lyapunov}
\end{equation}
The stability exponents $\gamma_p$ correspond to the Lyapunov exponents in
periodic orbit theory.

When only one eigenvalue of the classical evolution operator $U$ is on the
unit circle, one has, $u_n\overrightarrow {\scriptstyle {n\rightarrow \infty
}}\ 1$.  This leads to a classical sum-rule
\begin{equation}
u_n=\sum_{p\in P_n}n_p\exp (-r \gamma_p n_p ) \ \
\overrightarrow{
\scriptstyle {n\rightarrow \infty }}\ 1\,.
\label{classicalsum}
\end{equation}
This last relation shows again that the number of periodic orbits must
increase exponentially with $n$ to balance the exponentially decreasing
stability factors of the individual periodic orbits. The topological entropy
can be related to the mean stability exponent through this relation.

Using the expression (\ref{classicalsum1}) for $u_n$ one can easily write down
the complete thermodynamic formalism for the graph. Here, we shall only quote
the periodic orbit expression for the Ruelle $\zeta $ function
\begin{eqnarray}
\zeta _R(z) &\equiv &\left( \det (I-zU)\right) ^{-1}={\rm \exp }\left[ -{\rm
tr}\left( \ln (I-zU)\right) \right]  \label{cl4} \\
&=&\exp \left[ \sum_n\frac{z^n}nu_n\right] =\prod_p\frac 1{\left(
1-z^{n_p}\exp (-n_p\gamma _p)\right)}\,,  \nonumber
\end{eqnarray}
where the product extends over all primitive periodic orbits.

The above discussion of the classical dynamics on the graph shows that it
bears a striking similarity to the dynamics induced by area preserving
hyperbolic maps. The reason underlying this similarity is that even though the
graph is a genuinely one-dimensional system, it is not simply connected, and
the complex connectivity is the origin and reason for the classically chaotic
dynamics.
\section{The Spectral Statistics of the Quantum Scattering Map}
\label {statistics}
We consider the matrices $S_B$ defined in (\ref {S_Bdef}).  Their spectrum
consist of $2B$ points confined to the unit circle (eigenphases).  Unitary
matrices of this type are frequently studied since they are the quantum
analogues of classical, area preserving maps. Their spectral fluctuations
depend on the nature of the underlying classical dynamics \cite{S89}.  The
quantum analogues of classically integrable maps display Poissonian statistics
while in the opposite case of classically chaotic maps, the statistics of
eigenphases conform quite accurately with the results of Dyson's random matrix
theory (RMT) for the {\it circular} ensembles.  The ensemble of unitary
matrices which will be used for the statistical study will be the set of
matrices $S_B(k)$ with $k$ in the range $|k-k_0| \le \Delta_k/2$. The interval
size $\Delta_k$ must be sufficiently small such that the vertex matrices do
not vary appreciably when $k$ scans this range of values.  Then the $k$
averaging can be performed with the vertex scattering matrices replaced by
their value at $k_0$.  When the vertex scattering matrices are derived from
Neumann or Dirichlet boundary conditions, the averaging interval is
unrestricted because the dimension of $S_B$ is independent of $k$. In any case
$\Delta_k$ must be much larger than the correlation length between the
matrices $S_B(k)$, which was estimated in \cite {KS99} to be inversely
proportional to the width of the distribution of the bond lengths. The
ensemble average with respect to $k$ will be denoted by
\begin{equation}
\label{ensaveS}
\left \langle \ \cdot \ \right \rangle _k \equiv \frac {1}{\Delta_k}
 \int_{k_0-\Delta_k/2}^{k_0+\Delta_k/2}\cdot
\,\, dk \,.
\end{equation}
Another way to generate an ensemble of matrices $S_{B}$ is to randomise the
length matrix $L$ or the magnetic vector potentials $A_{(i,j)}$, while the
connectivity (topology of the graph) is kept constant. In most cases, the
ensembles generated in this way will be equivalent. In the last section we
will also consider an {\em additional} average over the vertex scattering
matrices.

In the following subsections we compare statistical properties of the
eigenphases $\left\{ \theta _l(k)\right\} $ of $S_B$ with the predictions of
RMT \cite{M90} and with the results of periodic orbit theory for the spectral
fluctuations of quantised maps \cite{BS88}. The  statistical measure which
we shall investigate is the spectral form factor. Explicit expressions for this
quantity are given by RMT \cite{HKSSZ96}, and a semiclassical
discussion can be found in \cite{keatbog,UScorr,camb}.
\subsection{\bf The Form Factor}
\label{the_form_factor}
The matrix $S_B$ for a fixed value of $k$ is a unitary matrix with eigenvalues
$\e^{\i\theta_l(k)}$. The spectral density of the eigenphases reads
\begin{equation}
d(\theta;k )\equiv \sum_{l=1}^{2B}\delta (\theta -\theta
_l(k))=\frac{2B}{2\pi }+
\frac 1{2\pi }\sum_{n=1}^\infty\e^{-\i\theta n}{\rm tr}S^n_B(k) +{\rm c.c.}\,,
\label{sms1}
\end{equation}
where the first term on the r.h.s.\ is the smooth density
$\overline{d}=\frac{2B}{2\pi }$. The oscillatory part is a Fourier series with
the coefficients ${\rm tr}S^n_B(k) $.  This set of coefficients will play an
important r\^ole in the following.  Using the definitions (\ref{S_Bdef}) one can
expand ${\rm tr}S^n_B(k) $ directly as a sum over $n-$periodic orbits on
the graph
\begin{equation}
{\rm tr} S^n_B(k)
=\sum_{p\in {\cal P}_n}n_p{\cal A}_p^r{\rm e}^{i(kl_p+
\Phi_p)r}{\rm e}
^{i
\mu _p  r }\,,
\label{posum}
\end{equation}
where the sum is over the set ${\cal P}_n$ of primitive PO's whose period
$n_p$ is a divisor of $n$, with $r=n/n_p$. $l_p = \sum_{b \in p} L_{b}$ is the
length of the periodic orbit.  $\Phi_p = \sum_{b \in p}L_b A_b$ is the
``magnetic flux" through the orbit. If all the parameters $A_b$ have the same
absolute size $A$ we can write $\Phi_p = A b_p$, where $b_p$ is the directed
length of the orbit. $\mu _p$ is the phase accumulated from the vertex matrix
elements along the orbit, and it is the analogue of the Maslov index. For the
standard vertex matrices (\ref {smatrix}) $\mu_p/\pi$ gives the number of {\em
backscatterings} along $p$. The amplitudes ${\cal A}_p$ are given by
\begin{equation}
{\cal A}_p=\prod _{j=1}^{n_p} \left
|\sigma^{(i_j)}_{i_{j-1},i{j+1}}\right| \equiv
{\rm e}^{-{\frac{ \gamma
_p}2}n_p}\,,  \label{amplitude}
\end{equation}
where $i_{j}$ runs over the vertex indices of the periodic orbit, and $j$ is
understood ${\rm mod}\,n_{p}$. The Lyapunov exponent $\gamma_p$ was defined in
(\ref {lyapunov}). It should be mentioned that (\ref {posum}) is the building
block of the periodic orbit expression for the spectral density of the graph,
which can be obtained starting from the secular equation (\ref {secular}).
In the quantisation of classical area preserving maps similar expressions
appear as the leading semiclassical approximations. In the present context
(\ref {posum}) is an identity.

The two-point correlations are expressed in terms of the excess probability
density $R_2(r)$ of finding two phases at a distance $r$, where $r$ is
measured in units of the mean spacing ${ 2\pi \over 2B}$
\begin{equation}
R_2(r;k_0)={2\over 2\pi}\sum_{n=1}^\infty \cos
\left(
\frac{2\pi rn}{2B}\right) \frac 1{2B}\left\langle\left|
{\rm
tr}S_B^n\right|^2\right\rangle_k\,\, .
\label{sms3}
\end{equation}
The form factor
\begin{equation}\label{ff}
K(n/2B)={\frac 1{2B}}<|{\rm
tr}S_B^n|^2>_k 
\end{equation}
is the Fourier transform of $R_2(r,k_0)$.  For a Poisson spectrum, $K(n/2B)=1$
for all $n$.  RMT predicts that $K(n/2B)$, depends on the scaled time ${n/2B}$
only \cite{S89}, and explicit expressions for the orthogonal and the unitary
circular ensembles are known \cite{HKSSZ96}.

As was indicated above, if the vertex scattering matrices are chosen by
imposing Dirichlet boundary conditions on the vertices, the classical dynamics
is ``integrable". One expects therefore the spectral statistics to be
Poissonian,
\begin{equation}
K(n/2B)= 1\qquad
{\rm for\ all}\ n\ge 1
\,.
\end{equation}
For Dirichlet boundary conditions the vertex scattering matrices
(\ref{smatrix}) couple only time reversed bonds. $S_B$ is reduced to a block
diagonal form where each bond and its time reversed partner are coupled by a
$2\times 2$ matrix of the form
\begin{eqnarray}
S^{(b)}(k,A)
=\left (
{\begin{array}{ll}
0  &  \e^{\i(k+A)L_b} \\
\e^{\i(k-A)L_b } & 0
\end{array}}\right ) \,.
\end{eqnarray}
The spectrum of each block is the pair $\pm \e^{\i kL_b}$, with the
corresponding symmetric and antisymmetric eigenvectors $ {1\over \sqrt {2}}(1,
\pm1)$.  As a result, we get
\begin{equation}
K(n/2B)=1+(-1)^n \ \
\
{\rm for \ \ all} \ \ \ n\geq 1 \,.
\label{poissonform}
\end{equation}
This deviation from the expected Poissonian result is due to the fact that the
extra symmetry reduces the matrix $S_B$ further into the symmetric and
antisymmetric subspaces. The spectrum in each of them is Poissonian, but when
combined together, the fact that the eigenvalues in the two spectra differ
only by a sign leads to the anomaly (\ref {poissonform}).

\begin{figure}[htb]
\centerline{\psfig{figure=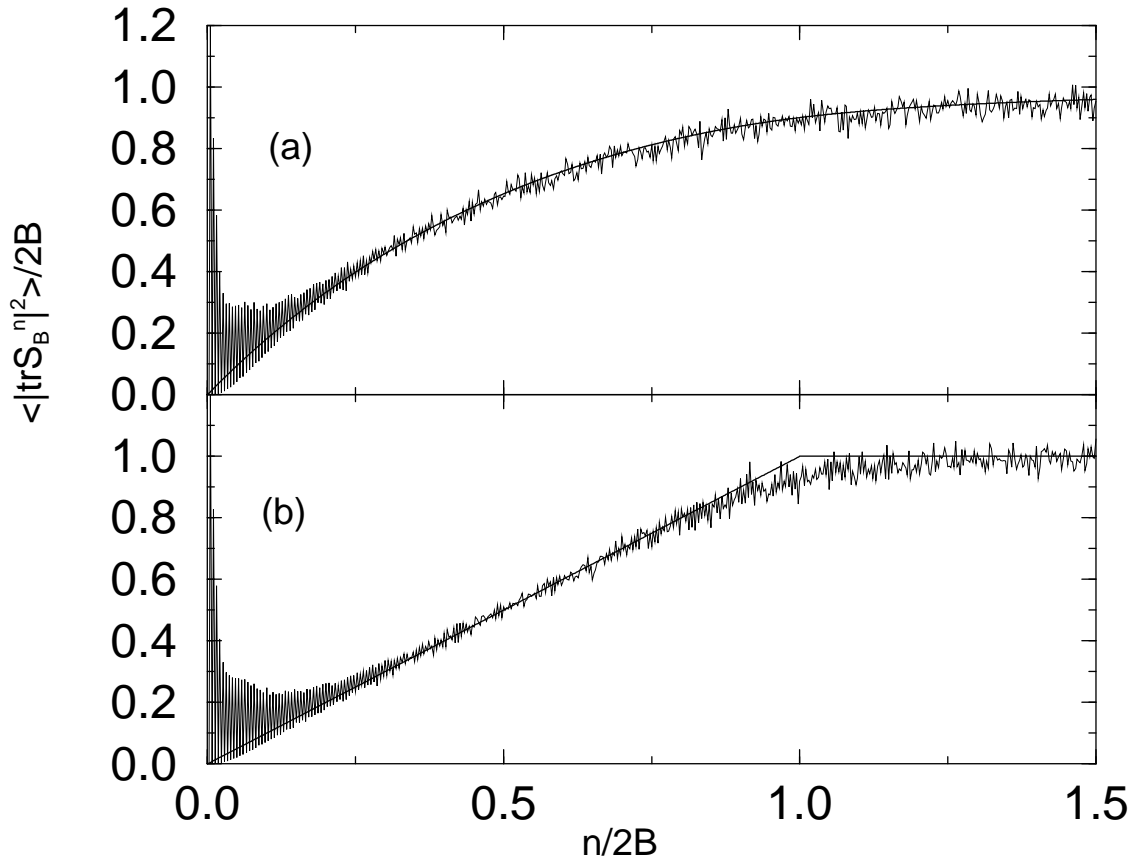,width=15
cm,angle=270}}
\vspace*{3mm}
\caption{\label{v20} Form factor for a fully connected graph with $V=20$ (a)
with and (b) without time-reversal symmetry. The smooth curves show the
predictions of the corresponding random matrix ensembles COE and CUE,
respectively.  }
\end{figure}

Having successfully disposed of the integrable case, we address now the more
general situation. In Fig.~\ref{v20} we show typical examples of form factors,
computed numerically for a fully connected graph with $V= 20$. The data for
Neumann boundary conditions and $A=0$ (Fig.~1(a)) or $A\ne 0$ (Fig.~1(b)) are
reproduced quite well by the predictions of RMT, which are shown by the smooth
lines. For this purpose, one has to scale the topological time $n$ by the
corresponding ``Heisenberg time" which is the dimension of the matrix, i.e.,
$2B$.  The deviations from the smooth curves are not statistical, and cannot
be ironed out by further averaging. Rather, they are due to the fact that the
graph is a dynamical system which cannot be described by RMT in all detail.
To study this point in depth we shall express the form factor in terms of the
PO expression (\ref{posum}).
\begin{eqnarray}
K(n/2B)&=&\frac 1{2B}\left\langle \left|\sum_{p\in {\cal P}_n}n_p{\cal
A}_p^r\e^{\i(kl_p+A
b_p+ \pi \mu_p)r}\right|^2 \right\rangle_k
\label{sms5} \\
&=&\left . \frac 1{2B} \sum_{p,p'\in {\cal P}_n}
n_pn_{p\prime} {\cal A}_p^r
{\cal A}_{p\prime}^{r^{\prime}}
\exp \left
\{\i A(r b_p-r'b_{p\prime}) +i\pi (r\mu_p-r'\mu_{p\prime})\right\}
 \right
|_{rl_p = r^{\prime}l_{p^{\prime}}}\,. \nonumber
\end{eqnarray}
The $k$ averaging is carried out on such a large interval that the double sum
above is restricted to pairs of periodic orbits which have exactly the same
length. The fact that we choose the lengths of the bonds to be rationally
independent will enter the considerations which follow in a crucial way.

The largest deviations between the numerical data and the predictions of RMT
occur for $n=1,2$.  For $n=1$ one gets $0$ instead of the COE (CUE) values
$1/B$ ($1/2B$), simply because the graph has no periodic orbits of period
$1$. This could be modified by allowing loops, which were excluded here from
the outset.  The $2$-periodic orbits are self-retracing (i.e.\ invariant under
time reversal), and each has a distinct length. Their contribution is enhanced
because back scattering is favoured when the valency is large.  Self-retracing
implies also that their contribution is insensitive to the value of $A$. The
form factor for $n=2$ calculated for a fully connected graph with $v=V-1$ is
\begin{equation}
\label{tsampi}
K(n/2B)= 2\(\[1-\frac 2v\]\)^4\,,
\end{equation}
independent of the value of $A$.  This is different from the value expected
from RMT.  The repetitions of the 2-periodic orbits are also the reason for
the odd-even staggering which is seen for low values of $\tau\equiv n/2B$.
They contribute a term which is $\approx 2\exp(-2V\tau)$ and thus decays
faster with the scaled time $\tau$ when the graph increases.

The deviations between the predictions of RMT and periodic orbit theory for
low values of $\tau$ are typical and express the fact that for deterministic
systems in general, the short time dynamics is not fully chaotic. The short
time domain becomes less prominent as $B$ becomes larger because the time $n$
has to be scaled by $2B$. This limit is the analogue of the limit $\hbar
\rightarrow 0$ in a general system.

Consider now the domain $2 <n \ll 2B$. The PO's are mostly of the irreducible
type, and the length restriction limits the sum to pairs of orbits which are
conjugate under time reversal. Neglecting the contributions from repetitions
and from self-retracing orbits we get
\begin{equation}
\label{transuzy1}
K(n/2B)\approx \frac 1{2B} \sum_{p\in {\cal P}_n}  n^2 {\cal A}_p^2 \ \
4\cos ^2A b_p  = {2n\over 2B} u_n \left \langle \cos ^2  A b_p\right
\rangle _n \,.
\end{equation}
The classical return probability $u_n$ approaches $1$ as $n$ increases (see
(\ref{classicalsum})).  Neglecting the short time deviations, we can replace
$u_n$ by $1$, and we see that the remaining expression is the classical
expectation of $\cos ^2 A b_p$ over PO's of length $n$. For $A=0$ this factor
is identically $1$ and one obtains the leading term of the COE expression for
$n\ll 2B$. If $A$ is sufficiently large $ \left \langle \cos ^2 A b_p \right
\rangle_n \approx 1/2 $, one obtains the short-time limit of the CUE
result. The transition between the two extreme situations is well described by
\begin{equation}
\label{transuzy2}
 \left \langle \cos ^2  A b_p\right \rangle _n \approx {1\over 2} \left (
 \e^{- A^2\left \langle L_b^2\right \rangle {n\over 2}} +1 \right ) \,.
\end{equation}
This formula is derived by assuming that the total directed length $b_p$ of a
periodic orbit is a sum of elementary lengths with random signs.

The basic approximation so far was to neglect the interference between
contributions of periodic orbits with different codes (up to time
reversal). This can be justified as long as periodic orbits with different
codes have different lengths. This is the case for low values of $n$. As $n$
approaches $B$ the degeneracy of the length spectrum increases, and for $n>2B$
all the orbits are degenerate. In other words, the restriction $rl_p =
r^{\prime}l_{p^{\prime}}$ in (\ref {sms5}) does not pick up a unique orbit and
its time reversed partner, but rather a group of {\em isometric} but distinct
orbits. Therefore, the interference of the contributions from these orbits
must be calculated. The relative sign of the terms is determined by the
``Maslov" index.  The computation of the interfering contributions from
different periodic orbits with neighbouring actions is an endemic problem in
the semiclassical theory of spectral statistics. These contributions are
referred to as the {\it non-diagonal} terms, and they are treated by invoking
the concept of periodic orbit correlations \cite {ADDKKSS93,CPS98}.  The
dynamical origin of these correlations is not known. In the case of graphs,
they appear as correlations of the ``Maslov" signs within a class of isometric
$n$-periodic orbits.

To compute $K(n/2B)$ from (\ref {sms5}) one has to sum the contributions of
all the $n$-periodic orbits after grouping together those which have exactly
the same lengths. We shall discuss the case $A=0$, so a further restriction on
the orbits to have the same directed length is not required here. Since the
lengths of the individual bonds are assumed to be rationally independent, a
group of isometric $n$-periodic orbits is identified by the non-negative
integers $q_i, i=1,\dots,B$ such that
\begin{equation} l_{\bf q} \equiv
\sum_{i=1} ^B q_i l_i
\qquad{\rm
with}\qquad\sum_{i=1}^Bq_i=n\,,
\label{qdef}
\end{equation}
i.e., each bond $i$ is traversed $q_i$ times. The orbits in the group differ
only in the {\it order} by which the bonds are traversed. We shall denote the
number of isometric periodic orbits by $D_{n}(\bf q)$. Note that not all the
integer vectors ${\bf q}$ which satisfy (\ref {qdef}) correspond to periodic
orbits. Rather, the connectivity required by the concept of an orbit imposes
restrictions, which render the problem of computing $D_n({\bf q})$ a very hard
combinatorial problem \cite {Urigavish}. Writing (\ref {sms5}) explicitly for
the case of a fully connected graph with Neumann vertex scattering matrices,
we get
\begin{equation}
K(n/2B)={1\over 2B}\left({2\over v}\right)^{2n}
\sum_{\bf q} \left|\sum_{\alpha=1}^{D_n(\bf q)} {n \over r_{\alpha}}
(- \xi )^{\mu_{\alpha}}\right| ^2 \ , \ \
{\rm with}  \ \ \ \xi \equiv \left({v-2\over 2}\right)\,,
\label {tracecomp}
\end{equation}
and the $\alpha$ summation extends over the $n$-periodic orbits in the class
${\bf q}$.  $\mu_{\alpha}$ is the number of back scattering along the orbit,
and $r_{\alpha}$ is different from unity if the orbit is a repetition of a
shorter primitive orbit of period $n/r_{\alpha}$.

Equation (\ref {tracecomp}) is the starting point of the new approach to
spectral statistics, which we would like to develop in the present paper. The
actual computation of (\ref {tracecomp}) can be considered as a {\em
combinatorial} problem, since it involves counting of loops on a graph, and
adding them with appropriate (signed) weights. For Neumann boundary
conditions, the weights are entirely determined by the connectivity of the
graph. Our numerical data convincingly show that in the limit of large $B$ the
form factors for sufficiently connected graphs reproduce the results of
RMT. The question is, if this relation can be derived using asymptotic
combinatorial theory.  The answer is not yet known, but we would like to show
in the next section that for a very simple graph one can use combinatorics to
evaluate the periodic orbit sums, and recover in this way the exact values of
the form factor.
\section{\bf The $2$-star Model}
In this section we will investigate the classical and quantum dynamics in a
very simple graph using two different methods.  We shall use periodic orbit
theory to reduce the computation of the trace of the classical evolution
operator $u_{n}$ and the spectral form factor $K(n/2B)$ to combinatorial
problems, namely sums over products of binomial coefficients. The result will
be compared to a straight forward computation starting from the eigenvalues of
the classical and quantum scattering maps.

An $n$-star graph consists of a ``central" vertex (with vertex index $o$) out
of which emerge $n$ bonds, all terminating at vertices (with indices
$j=1,\dots, n$) with valencies $v_j=1$. The bond lengths are $L_{oj}\equiv
L_j$. This simple model (sometimes called a {\it hydra}) was studied at some
length in \cite {KS99}.  The star with $n=2$ is not completely trivial if the
central vertex scattering matrix is chosen as
\begin {equation}
\sigma^{(o)}(\eta ) = \left (
{\begin{array} {ll}
 \cos \eta        & {\rm i}\sin \eta \\
 {\rm i}\sin \eta  & \cos \eta
\end{array}}  \right )\,,
\label {2-starsigma}
\end{equation}
where the value $0\le \eta \le \pi /2 $ is still to be fixed. The scattering
matrices at the two other vertices are taken to be $1$ and correspond to
Neumann boundary conditions.  The dimension of $U$ and $S_B$ is $4$, but it
can be immediately reduced to $2$: due to the trivial scattering at the
reflecting tips, $a_{jo}=a_{oj}\equiv a_j$ for $j=1,2$. In this representation
the space is labelled by the indices of the two loops (of lengths $2L_1$ and
$2L_2$ respectively) which start and end at the central vertex.  After this
simplification the matrix $S_B$ reads
\begin {equation}
S_B(k;\eta) =
\left (
{\begin{array} {ll}
    {\rm e}^{2ikL_1} & 0   \\
  0    &  {\rm e}^{2ikL_2}
\end{array}}  \right )
\left (
{\begin{array} {ll}
 \cos \eta        & {\rm i}\sin \eta \\
 {\rm i}\sin \eta  & \cos \eta
\end{array}}  \right )\,.
\label {2-starSB}
\end{equation}
We shall compute the form-factor for two ensembles. The first is defined by a
fixed value of $\eta = \pi/4$, and the average is over an infinitely large $k$
range. The second ensemble includes an additional averaging over the parameter
$\eta$. We will show that the measure for the integration over $\eta$ can be
chosen such that the model yields the CUE form factor. This is surprising at
first sight, since the model defined above is clearly time-reversal invariant.
However, if we replace $kL_1$ and $kL_2$ in (\ref{2-starSB}) by $L(k\pm A)$,
(\ref{2-starSB}) will allow for an interpretation as the quantum scattering
map of a graph with a single loop of length $L$ and a vector potential $A$,
i.e., of a system with broken time-reversal invariance (see Fig.~\ref{tst}). In
particular, the form factors of the two systems will coincide exactly, when an
ensemble average over $L$ is performed. Clearly, this is a very special
feature of the model considered, and we will not discuss it here in more
detail.
\subsection {\bf Periodic Orbit Representation of $u_n$}
The classical evolution operator corresponding to (\ref{2-starSB}) is
 \begin {equation}
U(\eta) = \left (
{\begin{array} {ll}
 \cos ^2\eta    & \sin ^2 \eta \\
 \sin ^2\eta    & \cos ^2 \eta
\end{array}}  \right )\,.
\label {2-starclass}
\end{equation}
The spectrum of $U$ consists of $\{1,\cos 2\eta \}$, such that
\begin{equation}\label{un}
u_n (\eta )=1+\cos ^n 2\eta\,.
\end{equation}
We will now show how this result can be obtained from a sum over the periodic
orbits of the system, grouped into classes of isometric orbits. This grouping
is not really necessary for a classical calculation, but we would like to
stress the analogy to the quantum case considered below.

The periodic orbits are uniquely encoded by the loop indices, such that each
$n$-tuple of two symbols $1$ and $2$ corresponds (up to a cyclic permutation)
to a single periodic orbit. When $n$ is prime, the number of different
periodic orbits is $N_2(n)=2+(2^n-2)/n$, otherwise there are small
corrections due to the repetitions of shorter orbits. These corrections are
the reason why it is more convenient to represent a sum over periodic orbits
of length $n$ as a sum over all possible code words, though some of these code
words are related by a cyclic permutation and consequently denote the same
orbit. If we do so and moreover replace the stability factor of each orbit by
(\ref{lyapunov}), the periodic orbit expansion of the classical return
probability becomes
\begin{eqnarray}\label{un_po}
u_{n}&=&
\sum_{i_{1}=1,2}
\dots
\sum_{i_{n}=1,2}\prod_{j=1}^{n}
P_{i_{j}\rightarrow i_{j+1}}\,,
\end{eqnarray}
where $j$ is a cyclic variable such that $i_{n+1}\equiv i_{1}$.  In fact
(\ref{un_po}) can be obtained without any reference to periodic orbits if one
expands the intermediate matrix products contained in $u_{n}=\Tr U^{n}$ and
uses $P_{i_{j}\rightarrow i_{j+1}}=U_{i_{j},i_{j+1}}(\eta)$.

We will now order the terms in the multiple sum above according to the classes
of isometric orbits. In the present case a class is completely specified by
the integer $q\equiv q_1$ which counts the traversals of the loop $1$, i.e.,
the number of symbols $1$ in the code word. Each of the $q$ symbols $1$ in the
code is followed by an uninterrupted sequence of $t_{j}\ge 0$ symbols $2$
with the restriction that the total number of symbols $2$ is given by
\begin{equation}
\sum_{j=1}^{q}t_{j}=n-q\,.
\end{equation}
We conclude that each code word in a class $0<q<n$ which starts with a symbol
$i_{1}=1$ corresponds to an ordered partition of the number $n-q$ into $q$
non-negative integers, while the words starting with $i_{1}=2$ can be viewed
as partition of $q$ into $n-q$ summands.

To make this step very clear, consider the following example: All code words
of length $n=5$ in the class $q=2$ are $11222$, $12122$, $12212$, $12221$ and
$22211$, $22121$, $21221$, $22112$, $21212$, $21122$. The first four words
correspond to the partitions $0+3=1+2=2+1=3+0$ of $n-q=3$ into $q=2$ terms,
while the remaining $5$ words correspond to
$2=0+0+2=0+1+1=1+0+1=0+2+0=1+1+0=2+0+0$.

In the multiple products in (\ref{un_po}), a forward scattering along the
orbit is expressed by two different consecutive symbols $i_{j}\ne i_{j+1}$ in
the code and leads to a factor $\sin^2\eta$, while a back scattering
contributes a factor $\cos^2\eta$ . Since the sum is over periodic orbits, the
number of forward scatterings is always even and we denote it with $2\nu$. It
is then easy to see that $\nu$ corresponds to the number of positive terms in
the partitions introduced above, since each such term corresponds to an
uninterrupted sequence of symbols $2$ enclosed between two symbols $1$ or vice
versa and thus contributes two forward scatterings. For the codes starting
with a symbol $1$ there are ${q\choose \nu}$ ways to choose the $\nu$ positive
terms in the sum of $q$ terms, and there are ${n-q-1\choose \nu-1}$ ways to
decompose $n-q$ into $\nu$ {\em positive} summands. After similar reasoning for
the codes starting with the symbol $2$ we find for the periodic orbit
expansion of the classical return probability
\begin{eqnarray}\label{un_ex}
u_{n}(\eta)&=&2\cos^{2n}\eta+\sum_{q=1}^{n-1}\sum_{\nu}
\[{q\choose \nu}{n-q-1\choose \nu-1}+{n-q\choose \nu}{q-1\choose \nu-1}\]
\sin^{4\nu}\!\eta\,\cos^{2n-4\nu}\!\eta
\nonumber \\
&=&2\cos^{2n}\eta+\sum_{q=1}^{n-1}\sum_{\nu}{n\/\nu}
{q-1\choose \nu-1}{n-q-1\choose \nu-1}\sin^{4\nu}\!\eta\,\cos^{2n-4\nu}\!\eta\,
\nonumber \\
&=&2\sum_{\nu}{n\choose 2\nu}\sin^{4\nu}\!\eta\,\cos^{2n-4\nu}\!\eta
\nonumber \\
&=&(\cos^{2}\!\eta+\sin^{2}\!\eta)^{n}+(\cos^{2}\!\eta-\sin^{2}\!\eta)^{n}\,,
\end{eqnarray}
which is obviously equivalent to (\ref{un}). The summation limits for the
variable $\nu$ are implicit since all terms outside vanish due to the
properties of the binomial coefficients. In order to get to the third line we
have used the identity
\begin{equation}\label{ci1}
\sum_{q=1}^{n-1}{q-1\choose \nu-1}{n-q-1\choose \nu-1}=
{n-1\choose 2\nu-1}
={2\nu\over n}{n\choose 2\nu}\,.
\end{equation}
It can be derived by some straightforward variable substitutions from
\begin{equation}
\sum_{k=l}^{n-m}{k\choose l}{n-k\choose m}={n+1\choose l+m+1}\,.
\end{equation}
which, in turn, is found in the literature \cite{prudnikov}.
\subsection {\bf Quantum Mechanics: Spacing Distribution and Form Factor}\label{qm}
Starting from (\ref {2-starSB}), and writing the eigenvalues as ${\rm
e}^{ik(L_1+L_2)} {\rm e}^{\pm i\lambda/2}$, we get for $\lambda$, the
difference between the eigenphases,
\begin{equation}
\lambda = 2\,{\rm arcos}\[\cos\eta\,\cos k(L_1-L_2)\]\,.
\label {2-starlambda}
\end{equation}
For fixed $\eta$, the $k$ averaged spacing distribution (which is essentially
equivalent to $R_{2}(r)$ for the considered model) is given by
\begin{eqnarray}
\label{2-starspacing}
P(\theta ;\eta) &=& {1\over
\Delta_k} \int _{k_0-\Delta_k/2}^{k_0+\Delta_k/2} {\rm d}k
\
 \delta
\left (\theta - 2 {\rm arcos} \left [\cos \eta
\cos k(L_1-L_2) \right ]
\right  )  \nonumber \\ \nonumber \\
&=& \left \{
{\begin {array} {cl}
0 &
\qquad\cos(\theta/2)> |\cos \eta\,| \\ \\
{\displaystyle\sin(\theta/2)
\over \sqrt{\displaystyle\cos^2\eta-\cos^2(\theta/2)}}
&
\qquad\cos(\theta/2) < |\cos\eta\,|
\end {array}} \right.
\end{eqnarray}
We have assumed that $\theta$ is the smaller of the intervals between the two
eigenphases, i.e. $0\le \theta \le \pi$.

The spacings are excluded from a domain centered about $0$ $(\pi)$, i.e.,
they show very strong level repulsion. The distribution is square-root
singular at the limits of the allowed domain.

$P(\theta ;\eta)$ can be written as
\begin
{equation}
P(\theta;\eta ) =
{1\over 2\pi} +  {1\over \pi}\sum_{n=1}^\infty
\cos(n\theta)\,\(
{1\over 2}\left\langle\left|{\rm tr}S_B(\eta) ^n\right|^2\right\rangle_k -1\)\,,
\label{2-starPassum}
\end{equation}
and, by a Fourier transformation, we can compute the form factor
\begin{equation}
K_{2}(n;\eta)={1\over2} \left\langle\left|{\rm tr}S_B(\eta) ^n\right|^2\right\rangle\,. 
\end{equation}
In particular, for $\eta =\pi/4$ one finds
\begin{eqnarray}\label{K2PI4_UZY}
K_{2}(n;\pi/4)&=&1+{(-1)^{m+n}\over 2^{2m+1}}{2m\choose m} \\
&\approx& 1 + {(-1)^{m+n}\over 2\sqrt{\pi n }}\,.
\label{k2pi4_uzy_app}
\end{eqnarray}
Where $ m =[n/2]$ and $[\cdot]$ stands for the integer part. The slow
convergence of $K_{2}(n;\pi/4)$ to the asymptotic value $1$ is a consequence
of the singularity of $P(\theta;\pi /4 )$.

We now consider the ensemble for which the parameter $\eta$ is distributed
with the measure ${\rm d}\mu(\eta) = |\cos\eta \sin\eta |{\rm d}\eta$. The
{\em only} reason for the choice of this measure is that upon integrating
(\ref {2-starPassum}) one gets
\begin {equation}
P(\theta) = 2\sin^2(\theta/2)\,,
\label {2-starCUE}
\end{equation}
which coincides with the CUE result for $2\times 2$ matrices. A Fourier
transformation results in
\begin {equation}
K_2(n) = \left  \{ {\begin {array} {ll} {1\over 2} & {\rm for} \ \ n=1 \\
1 & {\rm for} \ \ n\ge 2 \end {array} } \right.\,.
\label {2-starK(n)CUE}
\end {equation}
The form factors (\ref{K2PI4_UZY}), (\ref{k2pi4_uzy_app}) and
(\ref{2-starK(n)CUE}) are displayed in Fig.~\ref{tst} below.
\subsection{Periodic Orbit Expansion of the Form Factor}\label{po}
As pointed out at the end of section \ref{the_form_factor}, the $k$-averaged
form factor can be expressed as a sum over classes of isometric periodic
orbits.  The analogue of (\ref{tracecomp}) for the 2-star is
\begin{equation}
K_2(n;\eta)={1\over 2}\sum_{q=0}^{n}
\left|\sum_{\alpha=1}^{D_{n}(q)}{n\over r_{\alpha}}
\i^{2\nu_{\alpha}}\sin^{2\nu_{\alpha}}\!\eta\cos^{n-2\nu_{\alpha}}\!\eta
\right|^{2}\,,
\end{equation}
where the number of forward and backward scatterings along the orbits are
$2\nu_\alpha$ and $\mu_{\alpha}=n-2\nu_{\alpha}$, respectively.  Again, it is
very inconvenient to work with the repetition number $r_{\alpha}$, and
consequently we replace---as in the derivation of (\ref{un_ex})---the sum over
orbits by a sum over all code words and use the analogy with the compositions
of integer numbers to obtain
\begin{eqnarray}\label{K2eta}
K_2(n;\eta)&=&
\cos^{2n}\!\eta+{n^2\over 2}\sum_{q=1}^{n-1}
\[\sum_{\nu}{(-1)^{\nu}\/\nu}{q-1\choose \nu-1}{n-q-1\choose \nu-1}
\sin^{2\nu}\!\eta\,\cos^{n-2\nu}\!\eta\]^{2}\,.
\end{eqnarray}
The inner sum over $\nu$ can be written in terms of Krawtchouk polynomials
\cite{kp1,kp2} as
\begin{eqnarray}\label{K2eta_K}
K_2(n;\eta)&=& \cos^{2n}\!\eta+{1\/2}\sum_{q=1}^{n-1}
{n-1\choose n-q}\cos^{2q}\!\eta\sin^{2(n-q)}\!\eta
\[{n\over q}P_{n-1,n-q}^{(\cos^{2}\!\eta,\sin^2\!\eta)}(q)\]^2\,,
\end{eqnarray}
and the Krawtchouk polynomials are defined as in \cite{kp1,kp2} by
\begin{eqnarray}\label{krawtchouk}
P_{N,k}^{(u,v)}(x)=\[{N\choose k}(uv)^{k}\]^{-1/2}\sum_{\nu=0}^{k}
(-1)^{k-\nu}{x\choose \nu}{N-x\choose k-\nu}u^{k-\nu}v^{\nu}\qquad
\(\begin{array}{l}0\le k \le N\cr u+v=1\end{array}\)\,.
\end{eqnarray}
These functions form a complete system of orthogonal polynomials of integer
$x$ with $0\le x\le N$. They have quite diverse applications ranging from the
theory of covering codes \cite {cohen} to the statistical mechanics of
polymers \cite{schulten}, and are studied extensively in the mathematical
literature \cite{kp1,kp2}. The same functions appear also as a building block
in our periodic orbit theory of Anderson localisation on graphs \cite {HSUS}.
Unfortunately, we were not able to reduce the above expression any further by
using the known sum-rules and asymptotic representations for Krawtchouk
polynomials.  The main obstacle stems from the fact that in our case the three
numbers $N,k,x$ in the definition (\ref{krawtchouk}) are constrained by
$N=k+x-1$.

\begin{figure}[tb]
\centerline{\psfig{figure=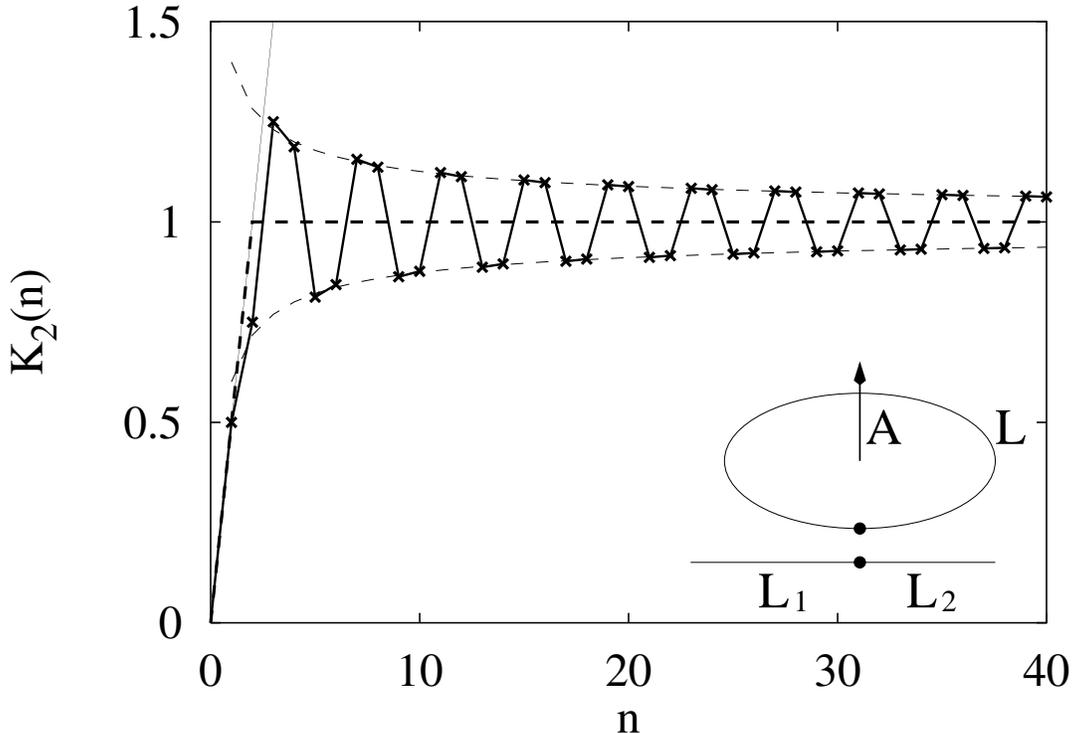,width=15cm,angle=270}}
\caption{\label{tst} Form factor for the 2-star quantum graph. The crosses and
the connecting heavy full line show the two equivalent exact results
(\protect\ref{K2PI4_UZY}) and (\protect\ref{k2pi4}) for $\eta=\pi/4$. The thin
dashed lines represent the approximation (\protect\ref{k2pi4_uzy_app}), and
the thin straight line corresponds to the diagonal approximation, when
repetitions of primitive periodic orbits are neglected. The heavy dashed line
exhibits the form factor of a CUE ensemble of $2\times 2$ random matrices
(\protect\ref{2-starK(n)CUE}), which can be obtained from the 2-star by an
appropriate averaging over $\eta$.  Finally, the inset shows a sketch of the
two possible realisations of the system: a time-reversal invariant 2-star
with bond lengths $L_{1}, L_{2}$ or a graph with a single loop of length $L$
and a magnetic flux $A$ breaking time-reversal symmetry.}
\end{figure}

We will now consider the special case $\eta=\pi/4$ for which we obtained in
the previous subsection the solution (\ref{K2PI4_UZY}). The result can be
expressed in terms of Krawtchouk polynomials with $u=v=1/2$ which is also the
most important case for the applications mentioned above. We adopt the common
practice to omit the superscript $(u,v)$ in this special case and find
\begin{eqnarray}\label{k2pi4}
K_2(n;\pi/4)&=&
{1\over 2^{n}}+{1\/2^{n+1}}\sum_{q=1}^{n-1}
{n-1\choose n-q}\[{n\over q}P_{n-1,n-q}(q)\right]^2\,.
\end{eqnarray}
It is convenient to introduce
\begin{eqnarray}\label{}
\nq(s,t)&=&(-1)^{s+t}{s+t-1\choose s}^{1/2}P_{s+t-1,s}(t)
\nonumber\\
&=&\sum_{\nu}(-1)^{t-\nu}{t\choose \nu}{s-1\choose \nu-1}
\end{eqnarray}
and to rewrite (\ref{k2pi4}) with the help of some standard transformations of
binomial coefficients as
\begin{eqnarray}\label{K2PI4N}
K_2(n;\pi/4)&=&{1\over 2^{n}}+{1\/2^{n+1}}\sum_{q=1}^{n-1}
\[{n\over q}\nq(q,n-q-1)\right]^2
\nonumber\\
&=&{1\over 2^{n}}+{1\over 2^{n+1}}\sum_{q=1}^{n-1}\[\nq(q,n-q)+(-1)^{n}\nq(n-q,q)\]^{2}
\end{eqnarray}
This expression is displayed in Fig.~\ref{tst} together with (\ref{K2PI4_UZY})
in order to illustrate the equivalence of the two results.  An independent
proof for this equivalence can be given by comparing the generating functions
of $K_2(n;\pi/4)$ in the two representations \cite{gregory}.  We defer this to
appendix \ref{proof}.
 
Please note, that in this way we have found a proof for two identities
involving Krawtchouk polynomials 
\begin{equation}\label{ci2a}
\sum_{q=1}^{2m-1}{2m-1\choose 2m-q}\[{2m\over q}P_{2m-1,2m-q}(q)\]^{2}
=2^{2m+1}+(-1)^{m}{2m\choose m}-2
\end{equation}
and
\begin{equation}\label{ci2b}
\sum_{q=1}^{2m}{2m\choose 2m+1-q}\[{2m+1\over q}P_{2m,2m+1-q}(q)\]^{2}
=2^{2m+2}-2\,(-1)^{m}{2m\choose m}-2\,,
\end{equation}
which were obtained by separating even and odd powers of $n$ in
(\ref{K2PI4_UZY}) and (\ref{k2pi4}). To the best of our knowledge,
(\ref{ci2a}) and (\ref{ci2b}) were derived here for the first time. 

Finally we will derive the CUE result (\ref{2-starK(n)CUE}) for the ensemble
of graphs defined in the previous subsection starting from the periodic orbit
expansion (\ref{K2eta}). We find
\begin{eqnarray}
K_{2}(n)&=&\int_{0}^{\pi/2}{\rm d}\mu(\eta) K_2(n;\eta)\,.
\end{eqnarray}
Inserting (\ref{K2eta}), expanding into a double sum and using
\begin{equation}
\int_{0}^{\pi/2}{\rm d}\eta
\sin^{2(\nu+\nu')+1}\!\eta\cos^{2(n-\nu-\nu')+1}\!\eta=
{1\over 2(n+1)}{n\choose \nu+\nu'}^{-1}
\end{equation}
we get
\begin{eqnarray}\label{intermediate}
K_{2}(n)&=&{1\/n+1}+
\\\nonumber &&+
{n^2\over 4(n+1)}\sum_{q=1}^{n-1}
\sum_{\nu,\nu'}
{(-1)^{\nu+\nu'}\/\nu\nu'}{n\choose \nu+\nu'}^{-1}
{q-1\choose \nu-1}{n-q-1\choose \nu-1}{q-1\choose \nu'-1}{n-q-1\choose
\nu'-1}\,.
\end{eqnarray}
Comparing this to the equivalent result (\ref{2-starK(n)CUE}) we were
again led to a previously unknown identity involving a multiple sum over
binomial coefficients. It can be expressed as
\begin{equation}\label{ci3}
S(n,q)=\sum_{\nu,\nu'}F_{\nu,\nu'}(n,q)=1\qquad (1\le q < n)
\end{equation}
with
\begin{eqnarray}
F_{\nu,\nu'}(n,q)&=&{(n-1)n\/2}{(-1)^{\nu+\nu'} \over \nu\nu'}
{n\choose \nu+\nu'}^{-1}
{q-1\choose \nu-1}{q-1\choose \nu'-1}{n-q-1\choose \nu-1}{n-q-1\choose
\nu'-1}\,.
\end{eqnarray}
In this case, an independent computer-generated proof was found \cite{akalu},
which is based on the recursion relation
\begin{equation}\label{recursion}
q^2F_{\nu,\nu'}(n,q)-(n-q-1)^{2}F_{\nu,\nu'}(n,q+1)+(n-1)(n-2q-1)F_{\nu,\nu'}(n+
1,q+1)=0\,.
\end{equation}
This recursion relation was obtained with the help of a Mathematica routine
\cite{multisum}, but it can be checked manually in a straight forward
calculation.  By summing (\ref{recursion}) over the indices $\nu,\nu'$, the
same recursion relation is shown to be valid for $S(n,q)$ \cite{multisum,A=B}
and the proof is completed by demonstrating the validity of (\ref{ci3}) for a
few initial values. Having proven (\ref{ci3}) we can use it to perform the
summation over $\nu,\nu'$ in (\ref{intermediate}) and find
\begin{eqnarray}
K_{2}(n)={1\/n+1}+\sum_{q=1}^{n-1}{n\/n^2-1}={1\/n+1}+{n\over
n+1}(1-\delta_{n,1})\,,
\end{eqnarray}
which is now obviously equivalent to the random matrix form factor
(\ref{2-starK(n)CUE}). To the best of our knowledge, this is the first
instance in which a combinatorial approach to random matrix theory is
employed.
\section{\bf Conclusions}
We have shown how within periodic orbit theory the problem of finding the form
factor (the spectral two-point correlation function) for a quantum graph can
be exactly reduced to a well-defined combinatorial problem. For this purpose
it was necessary to go beyond the diagonal approximation and to take into
account the correlations between the periodic orbits.

In our model, these correlations are restricted to groups of isometric
periodic orbits. This fits very well with the results of \cite{CPS98}, where
for a completely different system (the Sinai billiard), the classical
correlations between PO's were analysed and found to be restricted to
relatively small groups of orbits.  The code words of the orbits belonging to
one group were conjectured to be related by a permutation and a symmetry
operation, which is in complete analogy to the isometric orbits on graphs.

Even for the very small and simple graph model that we considered in the last
section the combinatorial problems involved were highly non-trivial. In fact
we encountered previously unknown identities which we could not have obtained
if it were not for the second independent method of computing the form factor.
However, since the pioneering work documented in \cite{A=B} the investigation
of sums of the type we encountered in this paper is a rapidly developing
subject, and it can be expected that finding identities like (\ref{ci2a}),
(\ref{ci2b}) and (\ref{ci3}) will shortly be a matter of computer power.

The universality of the correlations between periodic orbits in all chaotic
systems poses the problem to identify the common dynamical reasons for their
occurrence and to find a common mathematical structure which is capable to
describe them.  A very interesting question in this respect is, if the
correlations between PO's in a general chaotic system can be related to
combinatorial problems.
\section{\bf Acknowledgements}
This research was supported by the Minerva Center for Physics of Nonlinear
Systems, and by a grant from the Israel Science Foundation. We thank Tsampikos
Kottos for preparing the data for Fig.~\ref{v20}. We were introduced to the
{\it El Dorado} of combinatorial theory by Uri Gavish and we thank him as well
as Brendan McKay and Herbert Wilf for their interest and support. We are
indebted to Gregory Berkolaiko for his idea concerning the proof of
(\ref{ci2a}) and (\ref{ci2b}), and to Akalu Tefera for his kind help in
obtaining a computer-aided proof of (\ref{ci3}). HS wishes to thank the
Weizmann Institute of Science for the kind hospitality during the visit where
this work was initiated.
\appendix
\section{Proof of equivalence for Eqs.~(\protect\ref{K2PI4_UZY}) 
and (\protect\ref{K2PI4N})}\label{proof}
In this appendix we give an independent proof for the equivalence between the
two results (\ref{K2PI4_UZY}) and (\ref{K2PI4N}) obtained in sections
\ref{qm} and \ref{po}, respectively, for the form factor of the 2-star
with $\eta=\pi/4$. We define the generating function
\begin{eqnarray}\label{Gfun}
G(x)&=&\sum_{x=1}^{\infty}K_2(n;\pi/4)\,(2x)^{n}\qquad(|x|<1/2)
\end{eqnarray}
and find from (\ref{K2PI4_UZY})
\begin{eqnarray}\label{Gfun_uzy}
G(x)&=&{2x\/1-2x}-{1\over 2}+\sum_{m=0}^{\infty}{(-1)^{m}\/2}{2m\choose m}x^{2m}
(1-2x)
\nonumber\\
&=&{1\over 2}{1-2x\over \sqrt{1+4x^2}}-{1\over 2}{1-6x\/1-2x}\,.
\end{eqnarray}
On the other hand we have from (\ref{K2PI4N})
\begin{eqnarray}
G(x)={x\over 1-x}+G_{1}(x)+G_{2}(-x)
\end{eqnarray}
with
\begin{eqnarray}\label{G1}
G_{1}(x)=\sum_{s,t=1}^{\infty}\nq^{2}(s,t)\,x^{s+t}
\end{eqnarray}
and
\begin{eqnarray}\label{G2}
G_{2}(x)=\sum_{s,t=1}^{\infty}\nq(s,t)\,\nq(t,s)\,x^{s+t}\,.
\end{eqnarray}
A convenient starting point to obtain $G_{1}$ and $G_{2}$ is the integral representation
\begin{eqnarray}\label{irep}
\nq(s,t)=-{(-1)^{t}\over 2\pi\i}\oint{\rm d}z\,(1+z^{-1})^{t}(1-z)^{s-1}\,,
\end{eqnarray}
where the contour encircles the origin. With the help  of (\ref{irep}) 
we find
\begin{eqnarray}\label{g}
g(x,y)&=&\sum_{s,t=1}^{\infty}\nq(s,t)\,x^{s}\,y^{t}
\nonumber\\
&=&
-{1\over 2\pi\i}\sum_{s,t=1}^{\infty}
\oint{\rm d}z\,\sum_{s,t=1}^{\infty}(1+z^{-1})^{t}(1-z)^{s-1}\,x^{s}\,(-y)^{t}
\nonumber\\
&=&
{xy\over 2\pi\i}\sum_{s,t=0}^{\infty}\oint{\rm d}z\,
{1\over 1-x(1-z)}\,{1+z\over z+y(1+z)}
\nonumber\\
&=&
{xy\over (1+y)(1-x+y-2xy)}\qquad(|x|,|y|<1/\sqrt{2})\,.
\end{eqnarray}
The contour $|1+z^{-1}|=|1-z|=\sqrt{2}$ has been chosen such that both
geometric series converge everywhere on it. Now we have
{\small
\begin{eqnarray}
G_{1}(x^2)&=&{1\over (2\pi\i)^{2}}\oint{{\rm d}z\,{\rm d}z'\/zz'}
\sum_{s,t=1}^{\infty}\sum_{s',t'=1}^{\infty}\nq(s,t)\,\nq(s',t')\,
(x\,z)^{s}(x/z)^{s'}(x\,z')^{t}(x/z')^{t'}
\nonumber\\ 
&=&
{x^{4}\over (2\pi\i)^{2}}\oint{{\rm d}z\,{\rm d}z'}\,
{1\over (1+xz')(1+x[z'-z]-2x^2zz')}{z'\over (z'+x)(zz'+x[z-z']-2x^2)}\,,
\end{eqnarray}
}where $|x|<1/\sqrt{2}$ and the contour for $z,z'$ is the unit circle. We perform the double
integral using the residua inside the contour and obtain
\begin{eqnarray}
G_{1}(x)={x\over 2x-1}\({1\over \sqrt{4x^2+1}}-{1\/1-x}\)\,.
\end{eqnarray}
In complete analogy we find
\begin{eqnarray}
G_{2}(x)={1\over 2}{4x^2+2x+1\/(2x+1)\sqrt{4x^2+1}}-{1\over 2}
\end{eqnarray}
such that
\begin{equation}\label{GfunPO}
G(x)={x\/1-x}+{x\over 2x-1}\({1\over \sqrt{4x^2+1}}-{1\/1-x}\)+
{1\over 2}{4x^2-2x+1\/(1-2x)\sqrt{4x^2+1}}-{1\over 2}\,.
\end{equation}
The proof is completed by a straightforward verification of the equivalence
between the rational functions (\ref{Gfun_uzy}) and (\ref{GfunPO}).


\begin{thebibliography}{99}
\bibitem{KS97}  T.~Kottos and U.~Smilansky, Phys.\ Rev.\ Lett.\ {\bf 79},
4794 (1997).

\bibitem{KS99} T.~Kottos and U.~Smilansky, Annals of Physics, in press
(1999).

\bibitem{R83} Jean-Pierre Roth, in: {\it Lectures Notes in Mathematics:
Theorie du Potentiel}, A.~Dold and B.~Eckmann, eds., Springer, Berlin,
p.~521-539.

\bibitem{berry} M.~V.~Berry, {\it Proc. Royal Soc. London} {\bf A 400}, 229
(1985).

\bibitem{keatbog} E.~B.~Bogomolny and J.~P.~Keating, Phys.\ Rev.\ Lett.\ {\bf
77}, 1472 (1996).

\bibitem{ADDKKSS93} N.~Argaman, F. M. Dittes, E. Doron, S. P. Keating, A.
Y.~Kitaev, M.~Sieber, and U.~Smilansky, Phys.\ Rev.\ Lett.\ {\bf 71}, 4326 (1993).

\bibitem{CPS98} D.~Cohen, H.~Primack, and U.~Smilansky, Annals of Physics {\bf
264}, 108-170, (1998).

\bibitem{Miller97} D.~Miller, Phys.\ Rev.\ E {\bf 57}, 4063-4076 (1998).

\bibitem {Agam95} O.~Agam, B.~L. Altshuler, and A.~V. Andreev, Phys.\ Rev.\
Lett.\ {\bf 75}, 4389-4392 (1995).

\bibitem{A83}  S.~Alexander, Phys.\ Rev.\ B {\bf 27}, 1541 (1985).

\bibitem{A94} J.~E.~Avron, in: {\it Proc.~1994 Les Houches Summer School on
Mesoscopic Quantum Physics}, E.~Akkermans et al., eds., North-Holland,
p.~741-791.


\bibitem{Rochus} R.~Klesse and M.~Metzler, Phys. Rev. Lett. {\bf 79}, 721
(1997); R.~Klesse, Ph.~D. Thesis, Universit\"at zu K\"oln, AWOS-Verlag, Erfurt
(1996).

\bibitem{S89} U.~Smilansky, in {\em Proc.~1989 Les Houches Summer School on
Chaos and Quantum Physics}, M.-J.~Giannoni et al., eds., North-Holland,
p.~371-441.

\bibitem{HKSSZ96} F.~Haake, M.~Kus, H.-J.~Sommers, H.~Schomerus, and
K.~Zyckowski, J.~Phys.~A {\bf 29}, 3641 (1996).

\bibitem{Urigavish} U.~Gavish, private communication.

\bibitem{camb} U.~Smilansky, Semiclassical Quantisation of Maps and Spectral
Correlations, {\it Proc.~1997 of the NATO Advanced Study Institute
``Supersymmetry and Trace Formulae "}, I.~Lerner, ed., Cambridge (in press).

\bibitem{M90} M.~L.~Mehta, {\it Random Matrices and the Statistical Theory of
Energy Levels}, Academic, New York (1990); T.~A.~Brody, J.~Flores,
J.~B.~French, P.~A.~Mello, A.~Pandey, and S.~S.~M.~Wong, Rev.\ Mod.\ Phys.\ {\bf 53},
385 (1981).

\bibitem{BS88} R.~Bl\"umel and U.~Smilansky, Phys.\ Rev.\ Lett.\ {\bf 60}, 472
(1988); R.~Bl\"umel and U.~Smilansky, Phys.\ Rev.\ Lett.\ {\bf 64}, 241 (1990).

\bibitem {Feller} W.~Feller, {\em An introduction to Probability Theory and
Applications}, Jon Wiley and son, New York (1966).

\bibitem{UScorr} U.~Smilansky, Physica D {\bf 109}, 1767 (1997).

\bibitem{kp1} G.~Szeg\"o, {\em Orthogonal polynomials}, American Mathematical
Society Colloquium Publications, Vol.~23, New York (1959).

\bibitem{kp2} A.~F.~Nikiforov, S.~K.~Suslov, and V.~B.~Uvarov, {Classical
Orthogonal Polynomials of a Discrete Variable}, Springer Series in
Computational Physics, Berlin (1991).

\bibitem {cohen} G.~Cohen, I.~Honkala, S.~Litsyn, and A.~Lobstein, {\em
Covering Codes}, North Holland Mathematical Library, Vol.~54, (1997).

\bibitem {schulten}
K.~Schulten, Z.~Schulten, and A.~Szabo, Physica A {\bf 100}, 599-614 (1980).

\bibitem{HSUS} H.~Schanz and U.~Smilansky, in preparation.

\bibitem{prudnikov} A.~P.~Prudnikov; J.~A.~Bryckov; O.~I.~Maricev, {\em
Integrals and Series}, Vol.~1, Eq.~4.2.5.39, Gordon and Breach Science Publ.,
New York (1986). Note that in this edition the relation contains a misprint; the
correct form which we provided in the text can easily be proven using
\protect\cite{A=B}.

\bibitem{gregory} This idea was conveyed to us by G.~Berkolaiko.

\bibitem{multisum} K.~Wegschaider, {\em Computer Generated Proofs of Binomial
Multi-Sum Identities}, Diploma Thesis, RISC, J. Kepler University, Linz
(1997).

\bibitem{akalu} A.~Tefera, private communication.

\bibitem{A=B} M.~Petkov{$\check{\rm s}$}ek, H.~S.~Wilf, and D.~Zeilberger
{\em A=B},  AK Peters, Wellesley, Mass.\ (1996).
\end{thebibliography}
\end{document}